# Where Does MCP Security End? A Three-Tier Mitigation Mapping Framework for the OWASP MCP Top 10

Christopher Krakauer and Andreas Schaad

Offenburg University of Applied Sciences
Email: andreas.schaad@hs-offenburg.de

**Abstract.** The Model Context Protocol (MCP) has become a leading standard for connecting Large Language Models (LLMs) to external tools, software, and data sources. MCP architecture favors ease of integration over security, blending the data and control planes into a single natural language channel. This introduces confused deputy and prompt injection vulnerabilities that cannot be completely mitigated in the software development lifecycle (SDLC) of an MCP server. Structured security guidance for server developers has been outpaced by this adoption trend, leaving them without a clear way to identify where vulnerabilities can be addressed at the server layer and where architectural constraints create hard limits. This paper presents a framework based on OWASP MCP Top 10 mapped to three architectural tiers, developed using Design Science Research (DSR). Vulnerabilities are classified as server-primary, gateway-primary, or protocol-primary applying a cascading decision tree following four questions, resulting in reproducible tier assignments and confidence levels independent of who applies it. The framework was validated against an enterprise-grade Managed File Transfer (MFT) MCP server built in Go. Validation results showed that high-confidence server-tier controls worked as predicted, and low-confidence protocol-tier controls failed for architectural reasons beyond server implementation. The architectural ceiling encountered during practical validation is the main contribution and shows how secure-by-design development can reduce the MCP attack surface but not completely mitigate it.

**Keywords:** Model Context Protocol · MCP security · Agentic AI · OWASP · Secure-by-design · Threat modeling · Design Science Research

## 1 Introduction

Within less than two years, the Model Context Protocol has shifted from an internal integration convenience and AI-hobbyist tool to the de facto standard [1] for connecting LLM-based agents to external tools and data. The appeal is clear: it is a single client-server protocol over JSON-RPC 2.0 [2, 3] that replaces the per-vendor boilerplate, custom authentication flows, and varied schema formats that turned every tool integration into an engineering project. The reduction in integration cost that MCP enabled was a major driver of enterprise adoption [4, 5].

The design decisions that drove adoption are also the source of the security problem: to communicate easily with all LLMs, MCP must speak their language. The protocol carries data and control instructions on the same plane, in the same communication channel, where they are processed together inside the reasoning context of the LLM. Separating data from instructions is a security principle older than the personal computer, ignorance of which permitted the early phone-phreaking exploits more than fifty years ago [6], yet its violation is embedded in the fundamental architecture of MCP. This results in a host of vulnerabilities, with confused deputy behavior and indirect prompt injection at the forefront, that cannot be fully mitigated by a server developer, despite how carefully the code is written.

There are three clear questions left for the developers and architects building MCP servers: To what extent can secure-by-design development mitigate MCP risk at the server layer? Where must those controls meet external infrastructure? And where do they reach hard limits embedded in the protocol architecture? Existing work answers parts of these questions, but not all of them at once. Vulnerability taxonomies such as the OWASP MCP Top 10 [7] catalogue what is at risk, and scanning tools detect subsets of it, but neither maps vulnerabilities to the architectural layer where they can be directly addressed, nor expresses how reliably any control is expected to perform.

Contributions of this paper:

1. **A three-tier mitigation mapping framework** classifying each entry in OWASP MCP Top 10 by the architectural layer (server, gateway, or protocol) where it is primarily addressable.
2. **A four-question cascading decision tree** generating reproducible tier and confidence assignments independent of the analyst applying it.
3. **Graduated confidence levels** (High, Medium, Low) grounded in whether the underlying control is deterministic or subject to non-deterministic LLM behavior.
4. **The identification of an architectural ceiling** as a firm upper limit on achievable security confidence imposed by the blended data and control planes, independent of implementation quality.
5. **Empirical testing** through a participant-observer case study of an actual enterprise MFT MCP server built in Go, with documented negative findings included.

The framework is intentionally scoped to the perspective of a server developer. It proposes no changes to the MCP specification, and it does not evaluate defenses for the host or client that the server communicates with. The purpose of the framework is to identify what is within server developer control, what must be delegated, and what no developer can reach without a change to the protocol architecture.

The remainder of the paper is organized as follows. Section 2 introduces MCP, the core architectural flaw, and related work. Section 3 sets out the architectural properties that constitute the foundation of the framework. Section 4 sets out the Design Science Research method. Section 5 presents the framework, decision tree, confidence criteria, and complete mapping. Section 6 reports the case-study validation. Section 7 discusses the architectural ceiling, extensibility, comparison, and limitations. Section 8 concludes.

## 2 Background and Related Work

### 2.1 The Model Context Protocol

MCP offers a solution to the coordination problem created by fragmented tool integration. Connecting LLM applications to external tools previously required code that was vendor specific, used non-standard authentication, and had custom schema handling per-provider that had to be changed with each model or tool shift. MCP replaces this with a stateless client-server protocol built on JSON-RPC 2.0, allowing any MCP-enabled client to communicate with any MCP server through a shared interface protocol. [2, 3] Three primitives (tools, resources, and prompts) are exposed by the protocol, and it relies on a trust-delegation chain where access between the model and the connected servers is brokered by the host application. The protocol is built with multi-server workflows in mind, and multiple MCP servers are often seen in deployments.

This paper treats MCP adoption as established context. The relevant question is where that adoption led and what the architecture looks like on closer examination.

### 2.2 The Core Architectural Flaw

In secure systems, the data plane and the control plane are kept separate so that content is not reinterpreted as a command. This separation collapses with MCP. Tool descriptions, resource contents, and tool outputs are delivered into the shared context the LLM uses to weigh decisions. These are all expressed in the same natural language as the instructions that govern agent behavior. Because the model is unable to reliably distinguish data to be processed from an instruction to be followed, any data that enters the context window is potentially a control signal that can be used by threat actors.

This lays the foundation for the confused deputy problem in MCP. It allows for a server with legitimate authority to act against user interest through the content data it processes. Additionally, it is the reason why indirect prompt injection [8], where instructions are smuggled or obfuscated into tool responses or resource content, is better seen as a result of how the protocol routes information. How an LLM interprets the context it is given cannot be significantly changed through input sanitization at the server side in the many permutations natural language offers.

### 2.3 Related Work and the Gap

The security of the central LLM merits attention prior to moving on to MCP-specific vulnerabilities. The first systematic taxonomy for attacks via data channels ingested by models, instead of direct input from users, known as indirect prompt injection, was provided by Greshake et al. [9], and demonstrated that poisoned documents could be used to compromise retrieval-augmented applications. More than two years of security research have not shut the door that Greshake et al. had opened. Hackett et al. [10] tested and analyzed guardrail systems that had been deployed in response, where adversarial circumvention methods were demonstrably successful. Two paths run separately through LLM security literature: model behavior, what a model *is*, and model security, what happens when it *acts* through tool use on external systems. MCP merges them into one larger problem. Narajala and Narayan [11] published a threat model framework built for generative AI agents, covering prompt injection, privilege escalation, and supply chain threats for agentic workflows. Although it is broadly aimed at agentic AI instead of mapping mitigations to specific layers of MCP architecture, their contribution offers the nearest present framework to the one developed in this paper. The first example of the confused deputy problem, proposed by Hardy [12], was resolved with scoped authorization, a deterministic fix for a deterministic system. There is no equivalent fix for the confused deputy that is the LLM, whose responses to identical inputs can vary and whose reasoning cannot be inspected.

Zero-trust architecture [13] brings mature defense-in-depth and least-privilege principles, although several of its tenets are contradicted directly by MCP architecture. Data is trusted and enters the LLM reasoning context in an undifferentiated state from tool responses, resource content, and prompt templates, without access control at the resource level. Importantly, the probabilistic deputy the architecture is reliant on breaks the assumption of trust built into the zero-trust framework: there is no way an LLM-based deputy that interprets security instructions with a chance of failure greater than zero can satisfy these requirements. The enterprise perspective applied by Narajala and Habler [14] narrows the scope to enterprise-level MCP-specific concerns, providing a closer point of comparison for the framework presented here.

Literature on supply chain security provides well-developed defenses for threat exposure built around a code base. A common trait shared by traditional supply chain attacks is that there are forensic traces left behind: dependencies, build artifacts, and binaries house malicious code, and this trail can be targeted using Software Bill of Materials (SBOM) practices, tools for dependency auditing, and package integrity verification with cryptographic hashing. A supply chain that ignores this is built into the MCP ecosystem [15–17], opening an attack surface running on metadata that is natural language, and missing forensic tracing. A poisoned tool description is ingested into LLM reasoning context as tokens that cannot be statistically distinguished from legitimate instructions. When it successfully executes there is no binary artifact, build log, or persistent trace to inspect. The gap in defense between traditional supply chains and MCP cannot be addressed with existing tools because there is no support from the protocol for them to work with.

Specific research around MCP is emerging. The OWASP MCP Top 10 [7] is developed and validated by a community of security experts, cataloging the ten MCP-specific security issues and is the vulnerability taxonomy used for this paper. The most recent and granular taxonomy of MCP threats is offered by Shen et al. [18], covering 38 categories; despite the added granularity, the OWASP taxonomy was chosen because the mitigation mapping contribution is aimed at practical use for developers that have no need for an extensive threat catalogue. Scanning tools like MCP-Scan, MCPScan.ai, and Cisco's MCP Scanner [19–21] engage specific attack vectors using static analysis, dynamic flow detection, and rule-based signature matching. What is not offered by any of this is a structured map of the catalogued vulnerabilities to the architectural tier where each can be mitigated, matched with a confidence level for how reliably that mitigation holds. That is the gap addressed by this paper.

## 3 Architectural Sources of Risk

The three tiers of the framework are the only boundaries where a control can be placed. A mitigation can be deployed inside the server boundary or in the deployment infrastructure that sits between servers. When those options are exhausted, the protocol specification is what remains as the only option for eventual change. These layers exist as boundaries, independent of any vulnerability, and the server tier is the natural starting point. A control can be applied inside the server boundary, acting on a threat it can both reach and observe, written in code that can be tested and verified. What is determined by the framework is the routing. It helps decide at which layer to counter a specific threat, and how far a control placed there can be trusted. These are two distinct questions, answered by separate architectural boundaries. One governs how far a control can be trusted wherever it sits within the tiers, and the other two govern where a control must sit to reach and observe the threat. Each property below is one of these boundaries, and as they are passed, they move the mitigation and confidence further from the server-tier. The server tier is not mapped to any of these properties because that is the default starting point before any properties might be applied.

**Non-determinism (the limit on verification).** The key component that will ultimately act on MCP content data is a probabilistic LLM. Responses made by it to identical inputs can change, the reasoning is masked, and there is nothing within the architecture to distinguish an instruction token from a data token. As a result, controls given in the form of natural language have no deterministic enforcement. Because the same input might be handled with varying weight by the LLM across runs with slightly different context, unit testing faces a definitive challenge. This challenge goes deeper than probabilistic behavior of the LLM; there are no verification checks to determine if core reasoning and goals that drive the logic align with those of the user. In specific conditions models can actively hide them and work against security constraints, moving the subject from confusion into subversion. Demonstrated across major models that had passed standard safety testing [22, 23], this means the deputy is not only confused, opaque, and non-deterministic, but may hold ulterior motives. No level of output validation can identify outputs from models that have been created to be purposefully deceptive. This design influences how far a control can be trusted once placed. It draws the line between mitigations built on accepted cryptographic or policy techniques with reliable outcomes, and mitigations that rely on the LLM behaving predictably, which can be bypassed without any intrinsic defect in the control. It is the source of the confidence axis.

**The blended plane (the limit on reachability).** Information and instructions share a single channel and are combined within LLM context, resulting in all data being a potential attack vector. Controls built within the server can shape what enters the context and attempt to segregate and tag data, but once that content data is passed into model reasoning, there is no method for deployed code to reconstruct the data and instruction boundary it attempted to maintain, and which the protocol had no method to preserve. Threats rooted here are unreachable from any server or gateway, because the reasoning step that enables them sits with the LLM, below the level any deployed controls operate at. This is the architectural property that forces a tier no server or gateway developer can approach: the protocol tier, where root causes can only be resolved with direct change to the specification.

**The unverified supply chain and multi-server scope (the limit on observability).** There is no visibility from a single server into the other servers in a shared deployment stack that are sharing the same context. Contrary to traditional package supply chains, the MCP ecosystem offers no integrity checks when tool definitions are distributed, a common vector for indirect prompt injection via tool poisoning attacks. Forensic traces in dependencies, build artifacts, and binaries would be left in a traditional attack. An identified threat is reachable and can be addressed with standard deterministic controls, but supply chain observability is beyond the boundary of any one server. Cross-server telemetry, connection policy, and detection of shadow servers require a perspective above the individual server and outside the implementation scope of server developers. This property drives the control up, but not out of reach, into an infrastructure layer between servers and the protocol. That layer is the gateway tier.

**Synthesis.** Threats that a server can reach, observe, and address with a testable control remain at the server tier. Threats that can be reached and addressed deterministically, but fall beyond observability of an individual server progress to the gateway tier. Threats with their root cause linked to the blended plane and outside the scope of other layers move to the protocol tier, where non-determinism confirms that any remaining partial control is probabilistic and can be bypassed. The confidence axis is controlled similarly, with highest confidence where a control is verifiable and the threat is within the server boundary, and lowest where the blended plane and non-determinism overlap. These three boundaries are what the decision tree of Section 5 interrogates, one question at a time, to assign each vulnerability its tier and confidence.

## 4 Methodology

The work follows Design Science Research (DSR), which designs and evaluates a purposeful artifact to solve an identified problem rather than measuring an existing phenomenon [24, 25]. The artifact created here is the three-tier mapping framework. The DSR cycle is approached in the paper as follows: the problem is identified through architectural analysis (Sections 2–3); the solution objective is a framework that shows developers what is addressed in server design, what requires external controls, and what is unreachable without protocol change; the framework is designed in Section 5; it is demonstrated and validated against a real deployment in Section 6; and it is evaluated in Section 7.

For threat identification, STRIDE [26] was also used as a categorization overlay across the OWASP MCP Top 10. It offered an independent cross-reference on the taxonomy and surfaced one category (Denial of Service) that the OWASP set does not enumerate.

Validation uses a single participant-observer case study [27] with an enterprise MFT MCP server implemented in Go. The participant-observer role is a methodological choice with real consequences for evidential weight, addressed explicitly in Section 7.4. Two structural safeguards help constrain introduced bias, with the decision tree and evaluation dimensions being defined before being applied to the case. Negative findings are reported with the same treatment as successful ones.

No supervised experiments or statistical measurements are performed. The mapped framework is the contribution, validated in practice. There are no claims of absolute mitigation success. The framework identifies where controls belong and determines how reliably they hold.

## 5 The Three-Tier Mitigation Framework

### 5.1 Tier Definitions

Vulnerabilities are each assigned a primary tier, the layer where the identified root cause is most appropriately addressed. Secondary controls for defense-in-depth may exist at other layers but do not influence the primary assignment.

- **Server tier.** The tier within the scope and control of server developers. Deterministic controls at implementation level like input validation, scope-limited API enforcement, secure credential handling, integrity verification address it.

– **Gateway tier.** Reliant upon deployment infrastructure and associated budgeting. Controls are required outside the server boundary, including telemetry across servers, policy enforcement for connections, and anomaly detection.
– **Protocol tier.** Outside both server and gateway control. The root causes lie in the blended data and control plane and cannot be resolved without changes to the protocol specification.

An individual primary tier is designated for each OWASP MCP Top 10 entry, based on where mitigations can operate most effectively against the root cause of a vulnerability. When controls offer only partial coverage, secondary tiers are used, and this layered tiering is essential for defense-in-depth. A protocol-primary vulnerability such as MCP06 with server-layer sanitization applied is not entirely useless: it is still a partial control that can be implemented, but the framework recognizes and defines the limitations of that solution at the server level.

### 5.2 The Cascading Decision Tree

Tier and confidence assignments are produced by four cascading questions applied to each entry. The cascade is what makes the framework a usable instrument: reapplying it to the same entry returns the same result.

– **Q1 (Determinism):** *Can a unit test be written to prove this control works?* If not: the control is not deterministic and skip to Q3. If yes, continue to Q2.
– **Q2 (Bypass):** *Can the control be bypassed without exploiting an implementation bug?* If not, confidence level is High, the control is maintained within and beyond the server boundary. If yes, move to Q3.
– **Q3 (Boundary):** *Does the control behave predictably for threats that pass through the server but fail against threats originating outside the server scope?* If yes, confidence level is Medium. If not, move to Q4.
– **Q4 (Circumvention):** *Can this control be circumvented with encoding, obfuscation, novel phrasing, or exploiting LLM non-deterministic behavior?* If reached, the answer should be yes, resulting in Low confidence. If not, reassess decision tree application.

### 5.3 Confidence Levels

The questions in the tree should return one of the three confidence levels. Confidence is a more granular targeting than tier assignment and it depends on how much trust a correctly implemented control can warrant.

– **High, deterministic control.** Q1 and Q2 both pass. Mitigations rely on accepted standards like cryptographic, algorithmic, or policy-enforcement techniques. There are predictable and verifiable outcomes. Example: an OAuth validation flow with PKCE rejecting expired tokens.

- **Medium, bounded control.** When Q2 fails (or Q1 fails into Q3) and Q3 passes. The control is deterministic within the scope of the server boundary but the threat stretches beyond it as well. Example: logging at the server level is reliable for local processes, although it has no visibility or tracking of an attack chain across multiple servers.
- **Low, probabilistic control.** The architectural limit of the protocol is reached at Q4. Controls depend on filtering methods or non-deterministic assumptions by the LLM and can be bypassed. This gives the threat actor an architectural advantage. Example: Output sanitization used against prompt injection, where limited rule sets cannot keep up with permutation variability or the many ways natural language can be interpreted.

An important property with Low confidence levels is that they represent a structural limit, not a weak implementation. The control is circumvented by the system operating as designed, not by exploiting a defect.

### 5.4 Applying the Tree

The goal of the decision tree is to make tier and confidence assignment reproducible, independent of the operator. High-confidence entries like MCP01, MCP02, MCP07 pass both Q1 and Q2 and are quickly resolved due to the deterministic properties of their controls. The three traces below follow more complicated paths and are worked through to demonstrate how tier and confidence are reached differently.

For MCP04 (Software Supply Chain Attacks and Dependency Tampering), the integrity control is deterministic, where the checksum either matches or it does not, and the build fails. Q1 passes but Q2 does not, because the control can still be bypassed by a module compromised upstream that is never detected and passes verification. The threat operates outside the operational boundary of the server, confirmed by Q3, and is the main reason for keeping dependency trees as minimal as possible. The developer can verify integrity, but not safety. This split results in a Medium confidence level.

With MCP05 (Command Injection and Execution), Q1 passes because parameterized command construction can be tested. Q2 fails to pass because of a secondary indirect path, where threats can still be constructed following parameters executed by the LLM. Q3 passes and confirms that there are predictable controls in place, but not for threats from the LLM reasoning context, resulting in a Medium confidence level. The gap that remains is directly tied to the blended data and control plane of Section 2.2. The path for indirect injection is paved by the LLM not being able to differentiate normal instructions from context manipulation. This is an architectural condition, beyond the server layer, surfacing at the protocol tier. MCP05 carries Medium confidence for controls at the server layer, showing established input validation, parameter controls, and mature Go SDK libraries that are deterministic and testable, but there are no additional controls or observability into how reasoning context (via natural language) interprets controls coming from the server layer.

In MCP06 (Intent Flow Subversion) the controls at the server layer are output sanitization and input validation. Applying the decision tree to them arrives at Low confidence, where Q1 just passes, because sanitization can be unit tested against known injection patterns with predictable results. Q2 fails, because sanitization is bypassed without exploiting any inherent defect, which is a structural bypass demonstrated by Hackett et al. [10] through various forms of character injection and adversarial ML techniques that circumvent neural classifiers. Q3 also fails, and is where MCP06 differs from the Medium entries, which passed Q3 on predictable controls with clear boundaries where they were bypassed. Natural language variability overwhelms sanitization and the non-deterministic LLM can still interpret sanitized content as instructions, with sanitization being a control still within the operational scope of the server. Q4 confirms Low confidence, with controls that can be circumvented through encoding, obfuscation, novel phrasing, and non-determinism. There are no current controls or boundaries within the protocol schema for the architecture to enforce.

### 5.5 The Complete Mapping

The framework is presented as an application of the decision tree to all ten OWASP MCP Top 10 entries and is shown in Table 1. The STRIDE column is retained as a reference overlay. Confidence is not grouped by STRIDE category; it degrades along the architectural line. This suggests that architecture, rather than threat category, shapes where and how reliably a vulnerability can be mitigated.

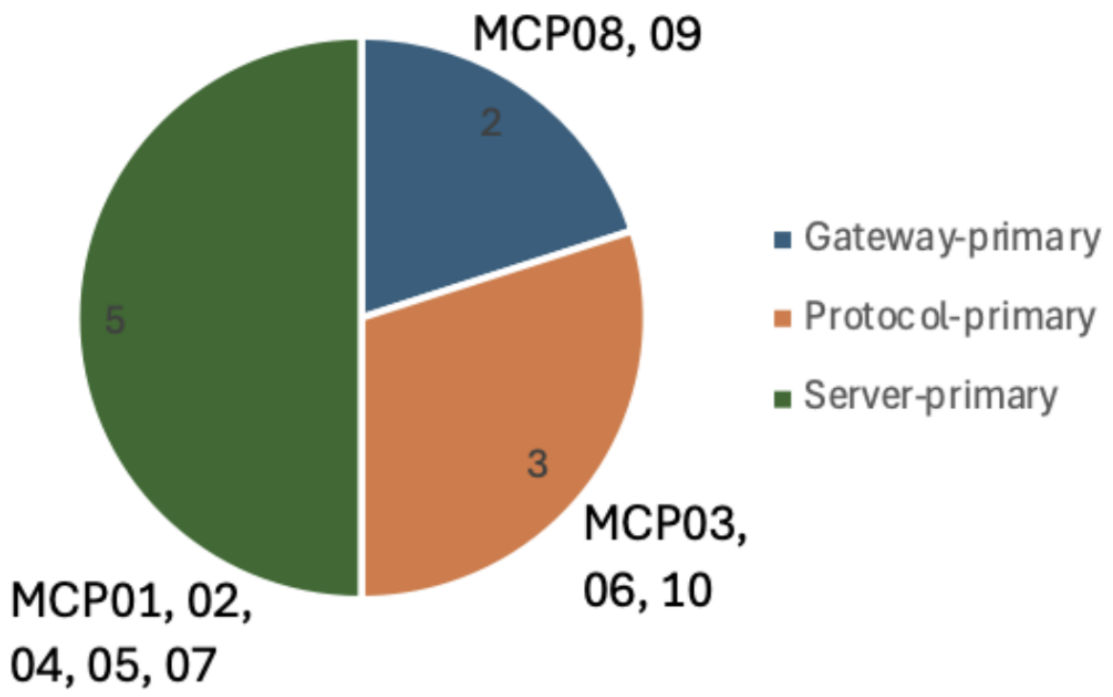

**Fig. 1.** Tier distribution, OWASP MCP Top 10.

### 5.6 Distribution Analysis

The distribution is not uniform, with the pattern linking directly to the architecture. There are five entries which are server-primary (MCP01, 02, 04, 05, 07), two exhibit as gateway-primary (MCP08, 09), and three are reached as protocol-primary (MCP03, 06, 10). By confidence, three entries are High (MCP01, 02, 07), four are Medium (MCP04, 05, 08, 09), and three show Low confidence (MCP03, 06, 10). Figure 1 shows the tier distribution.

**Table 1.** Decision tree applied to OWASP MCP Top 10.

| ID | Vulnerability | Tier | Conf. | Basis | STRIDE |
|---|---|---|---|---|---|
| MCP01 | Token Mismanagement & Secret Exposure | Server | High | Deterministic cryptographic controls. Threat within server boundary | I |
| MCP02 | Privilege Escalation via Scope Creep | Server | High | Deterministic policy enforcement. Threat within server boundary | E |
| MCP03 | Tool Poisoning | Protocol | Low | No primary control for attack vector. Cross-server descriptions bypass server visibility | T |
| MCP04 | Software Supply Chain Attacks & Dependency Tampering | Server | Medium | Deterministic integrity verification. Upstream threats beyond scope of server | T, R |
| MCP05 | Command Injection & Execution | Server | Medium | Direct path parameterized deterministic controls. Indirect LLM path bypasses | T, E |
| MCP06 | Intent Flow Subversion | Protocol | Low | All controls can be bypassed. Structural attack advantage with natural language | S, T, R, I, E |
| MCP07 | Insufficient Authentication & Authorization | Server | High | Deterministic established controls. Can fail, but requires architectural defect | S, E |
| MCP08 | Lack of Audit & Telemetry | Gateway | Medium | Reliable within server scope. Cross-server logging needs external aggregation | R |
| MCP09 | Shadow MCP Servers | Gateway | Medium | Reliable self-hardening. Neighboring servers outside operational boundary | S |
| MCP10 | Context Injection & Over-sharing | Protocol | Low | Server can mask some fields. LLM context aggregation probabilistic and outside server controls | I |

All entries with High confidence are reliant on a deterministic control with a threat contained within the server boundary, so Q1 and Q2 both pass. Each entry with Medium confidence shares the property that controls will work predictably within the scope of the server, but fail against threats beyond the observable boundary. The direction of the boundary failure cleanly separates the subgroups, with server-primary Medium entries facing threats leaking inward through LLM reasoning or upstream compromise, while gateway-primary Medium entries face threats moving outward across multi-server deployments. Entries with Low confidence do not split, with all being protocol-primary, being rooted in the blended plane, and reaching Q4.

## 6 Validation: An Enterprise MFT Case Study

For practical validation, the framework was applied to an enterprise Managed File Transfer (MFT) MCP server implemented in Go and developed under actual enterprise security governance. The deployment works as both a positive example of secure-by-design development and a controlled setting for locating the ceiling at the protocol layer. This shows that an established team applying a memory-safe language and strict governance will still run into the same structural limits shown in the framework.

### 6.1 Positive Validation: High-Confidence Controls Behave as Predicted

Entries with High confidence, found at the server tier, matched the results of the decision tree. Minimum API scopes are determined within the MFT software, and assigned upstream before the OAuth flow runs. This keeps access control inside reliable API infrastructure as seen with MCP02 and MCP07.

When observed, OAuth token validation with PKCE reliably rejected expired tokens, as seen with MCP01. In both cases, Q1 and Q2 passed, and observed behavior matched the High confidence level, showing that the control can be tested and cannot be bypassed without taking advantage of implementation defects. Figure 2 shows the confidence distribution across the taxonomy.

### 6.2 Negative Validation: Low-Confidence Entries Fail Structurally

Significant evidence was analyzed from documented failures, confirming assignments to the protocol tier by failing due to structural rather than implementation reasons. Failure of note:

- **Server-side trust tagging (UUID tagging)** [28]. A direct attempt to reintroduce data and instruction separation from the server layer, through wrapping content with provenance markers. It failed because nothing in the specification carries or recognizes such boundaries. The control was sound but the protocol offered no support for it (MCP06).

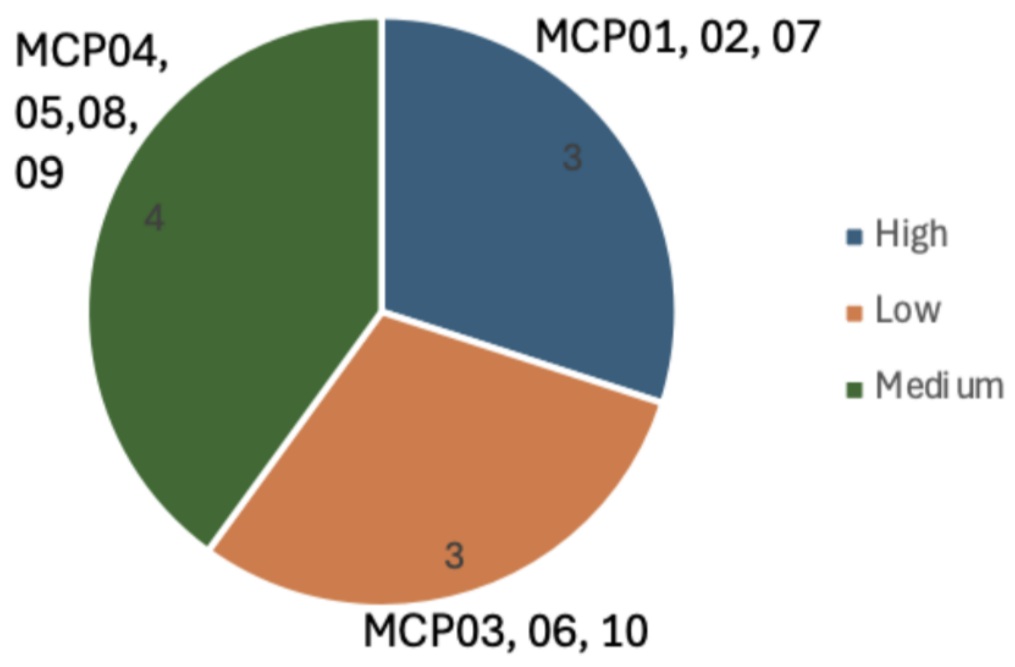

**Fig. 2.** Confidence distribution, OWASP MCP Top 10.

– **Field masking excluded.** Masking of sensitive fields was ultimately left out of the production deployment to preserve functionality for users, demonstrating the limits server controls face against context aggregation seen in MCP10.
– **Session isolation finds hard limits.** Attempts at session isolation ran into the same architectural limit, with per-server namespace being enforced within shared context.
– **Gateway controls deferred.** The enterprise MFT platform had no MCP-specific gateway controls, with them being deferred to customer discretion. This directly confirms that MCP08/MCP09 mitigation is an organizational or infrastructure responsibility rather than a server one.

These outcomes show that the tier and confidence assignments match constraints encountered in practice, with observations from failures providing significant data for analysis. Figure 3 summarizes confidence by tier.

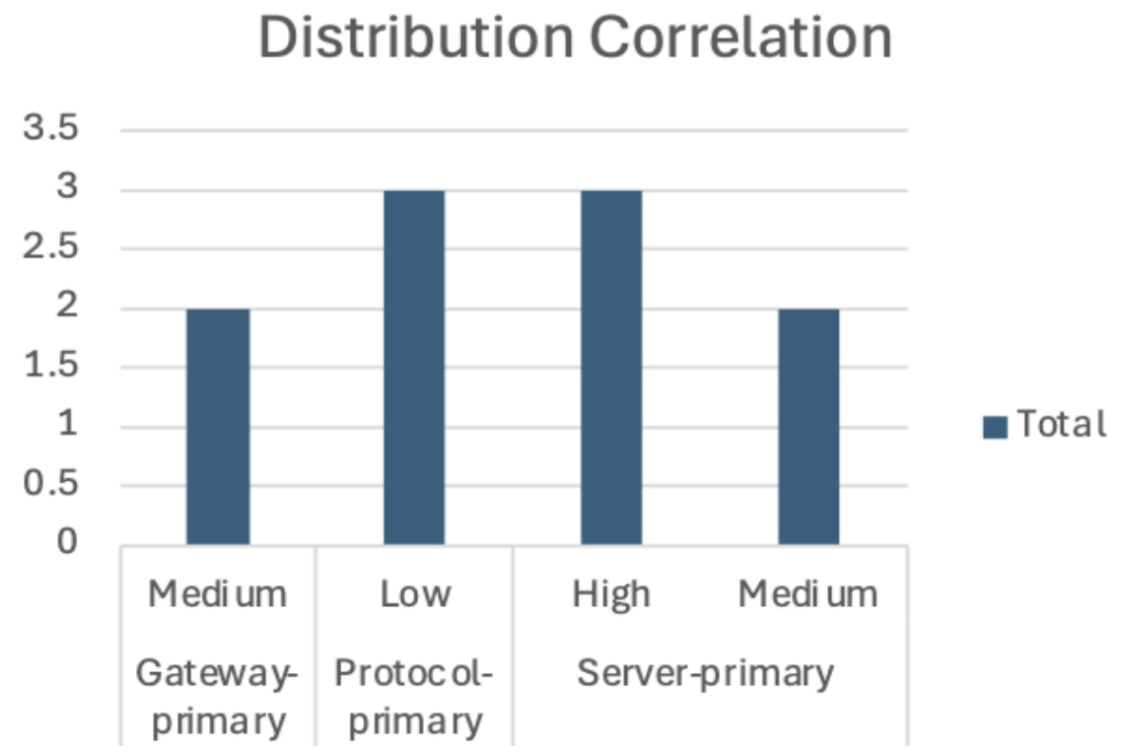

**Fig. 3.** Confidence distributed by tier.

# 7 Discussion

## 7.1 The Architectural Ceiling

The central finding is that all three entries that are protocol-primary with Low confidence share a single root cause: the blending of the data and control plane prevents any control within the server layer from deterministically separating data from instructions in the LLM context. A developer can try, but the boundary does not move with better implementation, a stronger SDK, or a more experienced team. This was evidenced by a capable enterprise team hitting the same wall, most clearly in the failed trust-tagging experiment. Secure-by-design development at the server layer is still necessary but not sufficient [29]. Half of the taxonomy can be addressed at the server, and neglecting it would leave a large attack surface unmitigated. Despite this, no level of server diligence can approach the protocol-tier entries, and attempting to develop mitigations for them can lead to a wrong turn that exhausts limited resources. This is the architectural ceiling, and locating it is the primary contribution of the framework.

## 7.2 Extensibility Beyond the Input Taxonomy

The framework is an instrument to be iterated through, extensible beyond the scope of the ten OWASP entries shown. A Denial of Service (DoS) vector can be dealt with through context-window saturation, which is absent from the OWASP taxonomy but present in STRIDE. In it, context space both in the finite computational form and as a monetary resource can be consumed and exhausted with accumulation of tool schemas and response payloads. This is listed by STRIDE as the Denial of Service category, although it has no equivalent entry in OWASP MCP Top 10.

The framework was never designed with a DoS entry in mind, which makes it an ideal test of extensibility. Even so, the three-tier structure and decision tree can process it without any changes. Controls at the server layer like rate limiting, response size caps, and tool timeout enforcement are all deterministic within the boundary of the server. Context window saturation as a DoS vector follows multiple paths, each leading to different points in the decision cascade.

**Control 1: Rate Limiting.** The direct control is rate limiting at the server layer. Q1 asks if a unit test can prove it works. A threshold test like this is proven and established. Q2 checks if the control can be bypassed without exploiting implementation bugs. As long as the attack path is through the server, it cannot be bypassed. Both questions pass. The confidence level is High for the direct path, and it is a deterministic control, so it should be at the server-tier beside MCP01, MCP02, and MCP07.

**Control 2: Aggregate Resource Monitoring.** There are threats beyond the server boundary, such as context window saturation. An individual server can cap personal response sizes, and it is testable. Q1 passes, but Q2 fails due to an attacker not requiring an implementation bug to exhaust context space if other servers beyond the control of the developer are feeding into the same context window. This moves to Q3, where it is asked if the control works predictably within server scope but fails to address threats from outside. Q3 passes because aggregate resource monitoring fulfills the requirement. The confidence level is Medium with a gateway-tier assignment. This parallels the decision tree pattern seen with MCP08 and MCP09.

**Control 3: Context Window Budgeting.** The last path is context window budget control. It enforces how much of the limited context space any single server can consume with output. There are no controls at the server layer and the server hits an observational boundary at the context window with no query parameters for remaining context space or protocol field to declare budgeting limits. Q1 fails because there is nothing at the server layer to unit test. There is no predictable control to assess because the contextual resource being exhausted is managed by the client and LLM. Q1 fails and skips to Q3. Q3 fails because of no predictable control. Q4 asks if the control can be circumvented through non-deterministic behavior of the system. In this case there is no control but the non-determinism is the effect. The probabilistic nature of the LLM governs how it allocates attention across accumulated context. The confidence level is Low, with a protocol-tier assignment, requiring changes to the specification to enable resource budgeting, with the entry being similar to MCP03, MCP06, and MCP10.

In this test for extensibility three control paths were explored through the same vulnerability, resulting in three different confidence levels, and tier assignments. This was reached following the same four questions without any modification to the decision tree or tier definitions.

### 7.3 Comparison with Current Approaches

Taxonomies are meant to catalogue, scanners aim to detect subsets, and gateways try to enforce policy. Outside of these, the framework holds a distinct position, offering a structured responsibility map showing developers where architectural responses can be made, and how reliably those solutions might work. Earlier approaches established what is vulnerable, with the framework adding which layer is responsible and how much confidence the response deserves.

### 7.4 Limitations

Evidence for this paper is drawn from a single case, a participant-observer role, applying qualitative rather than quantified confidence levels. This type of single-case validation limits the verified application scope. The prediction that tier assignments hold across deployments while confidence levels shift with SDK and dependency changes can be tested, but is untested here. The participant-observer role is recognized as a source of bias, with the decision tree and reporting of negative findings helping to act as constraints. Determination of confidence levels is reached through reproducible categorical judgments, where statistical quantification would require structured adversarial red-teaming against each entry. This would be in addition to the difficult task of establishing control benchmarks with a probabilistic system. These are all limits on evidential weight, not on the instrument: the decision tree still produces the same assignments independent of who applies it.

## 8 Conclusion

The attack surface of MCP servers is measurably reduced, though not eliminated, through secure-by-design development at the server layer. The framework presented structurally answers the three-part developer question and shows that building deterministic controls into the server layer can be used to approach half of the OWASP MCP Top 10 categories with High or Medium confidence. Additionally, two more entries depend on gateway infrastructure beyond any single MCP server, and the last three are limited to an architectural ceiling imposed by the blended data and control plane that is a feature of the protocol. A reproducible decision tree was applied to develop a complete tiered mapping framework, validated against a practical enterprise deployment, documenting both successful and failed controls. Together these led to the main contribution, and identification of the blended-plane architectural ceiling that recurred throughout, imposing a hard limit for developers. Future work shifts away from qualitative and into empirical confidence, with validation moving to different SDKs and deployment models. Red-teaming would be needed to convert the three confidence levels into measured bypass rates, with refinement of the decision tree and investigation into automation options for tier and confidence assignment flows to also be explored. Although beyond the immediate future work, a governing body for AI communication protocols, specifically involving tool use, seems to be the only direction where any major secure changes to the protocol might be made. MCP is currently lacking the kind of neutral, multi-stakeholder governance (IETF, W3C) [30, 31] under which critical security Internet protocols evolved.